\documentclass{ringb}
\usepackage{graphics}

\date{To appear in the Proceedings of the Ringberg Workshop on
{\it Diffuse Thermal and Relativistic Plasma in Galaxy Clusters}, editted
by Hans B\"ohringer (Garching: MPE Report).}

\begin{document}
\title{Models for the Relativistic Electron Population and Nonthermal Emission
in Clusters of Galaxies}
\author{Craig L. Sarazin}  
\institute{Department of Astronomy, University of Virginia,
P. O. Box 3818, Charlottesville, VA 22903-0818, U.S.A.}
\authorrunning{Craig L. Sarazin}
\titlerunning{Models for Relativistic Electrons and Nonthermal Emission in
Clusters}
\maketitle

\begin{abstract}
Models for the integrated relativistic electron population in clusters of
galaxies are presented.
The results depend on the history of electron acceleration in the cluster.
If there is no present particle acceleration or other sources, then no
high energy electrons ($\gamma \ga 300$, $E \ga 150$ MeV) should be present
due to inverse Compton (IC) losses.
High energy electrons are present if particle acceleration is occurring 
at present, perhaps as a result of a cluster merger shock.
The resulting IC, synchrotron, and gamma-ray emission from these models has
been calculated.
In the models, extreme ultraviolet (EUV) and very soft X-ray emission
are a nearly ubiquitous feature, because this emission comes from IC by
electrons with the longest lifetimes in clusters.
Diffuse UV and optical emission is also expected in most clusters, but
at levels which will be difficult to observe.
Hard X-ray tails and diffuse radio synchrotron emission are only expected
in clusters with recent or current particle acceleration;
for example, this acceleration might be due to intracluster merger shocks.
Gamma-ray observations at energies of around 100 MeV should detect both
the dominant population of relativistic electrons and the corresponding
ions.
The predicted fluxes are easily detectable with GLAST.
\end{abstract}

\section{Introduction} \label{sec:intro}

There are a number of reasons to think that cosmic ray particles
might be particularly abundant within the intracluster medium.
First, clusters of galaxies should be very effective traps for cosmic ray
ions and electrons.
Under reasonable assumptions for the diffusion coefficient, particles
with energies of less than $\la$$10^6$ GeV have diffusion times which
are longer than the Hubble time
(Colafrancesco \& Blasi 1999).
Second, the lifetimes of cosmic ray particles, even the electrons
which are responsible for most of the radiative signatures of relativistic
particles, can be quite long.
The radiation fields (optical/IR and X-ray) and magnetic fields
($B \la 1 \, \mu$G) in the ICM are low enough that high energy electrons
mainly lose energy by inverse Compton (IC) scattering of Cosmic Microwave
Background (CMB) photons
(Sarazin \& Lieu 1998).
Lower energy electrons can lose energy by Coulomb interactions with
the plasma;
however, at the very low densities ($n_e \la 10^{-3}$ cm$^{-3}$) in the
bulk of the ICM, this is only important for electrons with
$ \gamma \la 200$.
(Here, $\gamma$ is the Lorentz factor for the electrons, so that their
total energy is $\gamma m_e c^2$.)
For electrons with $\gamma \ga 200$, the lifetime is set by IC losses and
is
\begin{equation} \label{eq:lifetime}
t_{IC} = \frac{(\gamma - 1 ) m_e c^2}{\frac{4}{3} \sigma_T c \gamma^2
U_{CMB}}
\approx
7.7 \times 10^{9}
\left( \frac{\gamma}{300} \right)^{-1} \,
{\rm yr}
\, .
\end{equation}
The lifetimes of ions are set by interactions; for protons, this gives
$t_{ion} \ga 10^{11} ( n_e / 10^{-3} \, {\rm cm}^{-3})^{-1}$ yr.
Thus, clusters of galaxies can retain low energy electrons
($\gamma \sim 300$) and nearly all cosmic ray ions for a significant
fraction of a Hubble time.

Cluster of galaxies are likely to have substantial sources of cosmic rays.
Clusters often contain powerful radio galaxies, which may
produce and distribute cosmic rays throughout the cluster, in addition
to possible contributions from the cluster galaxies.
However, the primary reason why the cosmic ray populations in clusters
might be large is connected with the high temperature of the intracluster
gas.
This indicates that all of the intracluster medium
(typically, $10^{14} \, M_\odot$ of gas) has passed through strong
shocks with shock velocities of $\sim$1000 km/s during its history.
In our own Galaxy, whenever diffuse gas undergoes a strong shock
at velocities of this order, a portion of the shock energy
goes into the acceleration of relativistic particles
(e.g., Blandford \& Eichler 1987)
Thus, it seems likely that relatively efficient particle acceleration also
occurs in clusters of galaxies.

Direct evidence for the presence of an extensive population of
relativistic particles and magnetic fields in the ICM comes from
the observation of diffuse synchrotron radio halos in clusters
(e.g., Giovannini et al.\ 1993).
More recently, extreme ultraviolet (EUV) and very soft X-ray emission
has been detected from a number of clusters
(Lieu et al.\ 1996a,b;
Mittaz, Lieu, \& Lockman 1998).
One hypothesis is that this radiation is inverse Compton (IC) emission by
relativistic electrons
(Hwang 1997;
En{\ss}lin \& Biermann 1998;
Sarazin \& Lieu 1998).
Finally, hard X-ray emission has recently been detected in clusters with
$BeppoSAX$, which might be due to IC emission from higher energy electrons
(Fusco-Femiano et al.\ 1999;
Kaastra, Bleeker, \& Mewe 1998).
The EUV emission would require electrons with $\gamma \sim 300$, while
the hard X-ray emission would require $\gamma \sim 10^4$.

In this paper, illustrative models will be presented for
the integrated energy spectrum of primary, relativistic electrons,
under the assumption that they remain trapped in the cluster
(Colafrancesco \& Blasi 1999).
More details on these models are given in Sarazin (1999).
Radio synchrotron and inverse Compton emission spectra will
be determined up to the hard X-ray spectral band.
Gamma-ray emission by the same electrons (as well as emission by ions
through $\pi^o$ decay) will also be discussed briefly.

\section{The Electron Spectrum in Clusters} \label{sec:espect}

I will use the Lorentz factor $\gamma$ of the electrons as the
independent variable rather than the kinetic energy
$E = ( \gamma - 1 ) m_e c^2$, where $m_e$ is the electron mass.
Let $N( \gamma ) d \gamma$
be the total number of electrons in the cluster with
energies in the range $\gamma$ to $\gamma + d\gamma$, and let
$Q( \gamma ) d \gamma$ 
be the total rate of production of new cosmic ray this energy range.
The equation for the electron spectrum is then
\begin{equation} \label{eq:evolution}
\frac{\partial N( \gamma )}{\partial t} =
\frac{\partial}{\partial \gamma}
\left[ b( \gamma )  N( \gamma ) \right]
+ Q( \gamma )
\, ,
\end{equation}
where $b( \gamma ) = - d \gamma / d t$ gives the energy loss
by an individual electron.

\begin{figure}
\resizebox{\hsize}{!}{\rotatebox{-90}{\includegraphics{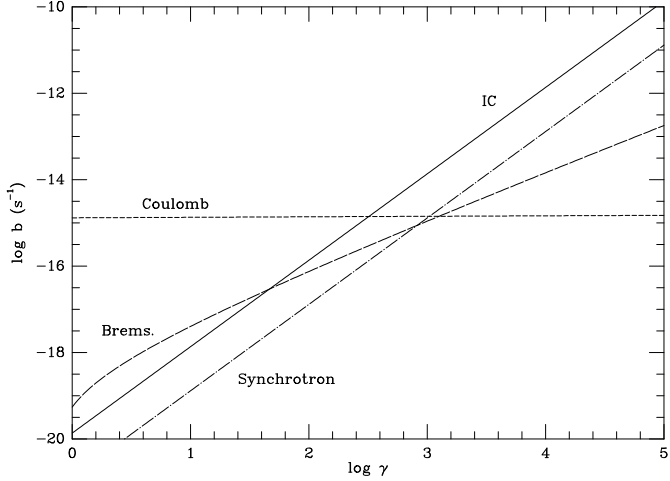}}}
\caption[]{The values of the losses function $b ( \gamma )$ for inverse
Compton (IC) emission, Coulomb losses, synchrotron losses, and bremsstrahlung
losses as a function of $\gamma$.
The values assume $n_e = 10^{-3}$ cm$^{-3}$, $B = 1 \, \mu$G, and
$z = 0$.
\label{fig:losses}}
\end{figure}

\begin{figure}
\resizebox{\hsize}{!}{\rotatebox{-90}{\includegraphics{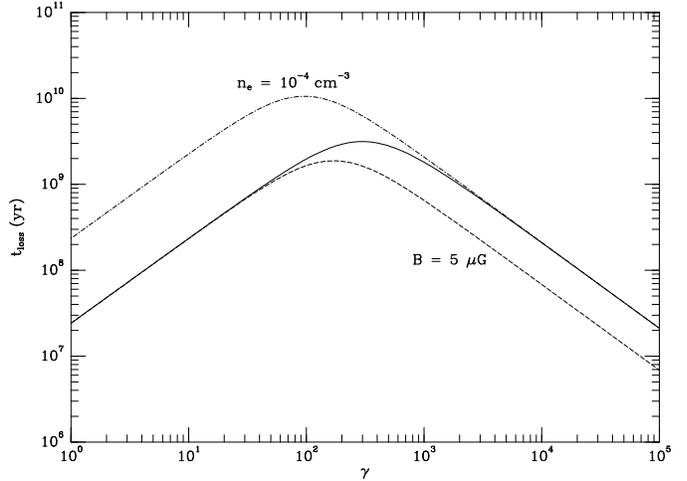}}}
\caption[]{The solid curve gives the loss time scale $t_{loss}$
as a function of $\gamma$ for electrons in a cluster with an electron density
of $n_e = 10^{-3}$ cm$^{-3}$ and a magnetic field of $B = 1$ $\mu$G.
The short-dash curve is for $B = 5$ $\mu$G, while the dash-dot curve
is for $n_e = 10^{-4}$ cm$^{-3}$.
\label{fig:lifetime}}
\end{figure}

The loss function includes IC emission,
Coulomb losses, synchrotron losses, and bremsstrahlung
losses.
In Fig.\,\ref{fig:losses},
these loss rates are shown as a function of $\gamma$ for typical
cluster conditions.
IC losses dominate at high energies $\gamma \ga 200$, while
Coulomb losses dominate at low energies.
Values for the loss time scale, defined as
$t_{loss} \equiv \gamma / b( \gamma )$,
are shown in Fig.\,\ref{fig:lifetime}.
The lifetime are maximum at $\gamma \approx 300$, where they
are $\sim 3 - 10$ Gyr, which is comparable to the likely ages of clusters.

\begin{figure}
\resizebox{\hsize}{!}{\rotatebox{-90}{\includegraphics{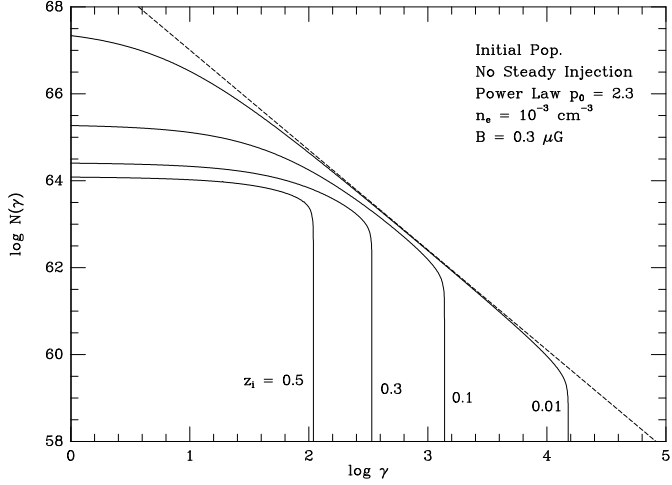}}}
\caption[]{The present day relativistic electron populations in models with
no current particle acceleration (e.g., no subcluster merger at present).
An initial population of electrons, which is shown as a dashed line, was
introduced into the cluster at a redshift of $z_i = 0.01$, 0.1, 0.3, and
0.5.
\label{fig:model_init}}
\end{figure}

The nature of the resulting electron populations depends on whether
there is a current or very recent ($z \la 0.03$) source of particles in the
clusters, such as an ongoing merger shock, or the electrons were all
accelerated in the past ($0.03 \la z \la 1$).
In all of the models shown here, the accelerated electrons have a power-law
spectrum, $Q( \gamma ) = Q_1 \gamma^{-p}$ with $p = 2.3$, and the
electron spectra are normalized such that the
total kinetic energy injected in electrons with $\gamma \ge 300$
is $E^{tot}_{CR,e} = 10^{63}$ ergs.
Fig.\,\ref{fig:model_init} shows the present-day electron spectra in
a series of models without any current source of newly accelerated
particles.
Thus, these models might apply to a cluster which is not currently
undergoing a subcluster merger.
At low energies, the electron population is reduced and the shape flattened
by Coulomb losses.
At high energies, the electron population is reduced and the shape of
the spectrum steepened by IC and synchrotron losses.
Because the IC and synchrotron losses increase with the square of the
electron energy,
there is an upper cut-off to the electron distribution;
no electrons are found at energies higher than $\gamma_{max}$.
Because the energy density in the CMB increases as $(1 + z)^4$, the
IC losses increase rapidly at high redshift.
If there is no electron acceleration since $z \ga 1$, the value of
$\gamma_{max} \la 100$, and Coulomb losses remove all of the lower
energy electrons.
In order to have any significant population of primary electrons with
$\gamma \ga 10^2$ at the present time,
there must have been a substantial injection of particles into clusters
at moderately low redshifts, $z \la 1$.

\begin{figure}
\resizebox{\hsize}{!}{\rotatebox{-90}{\includegraphics{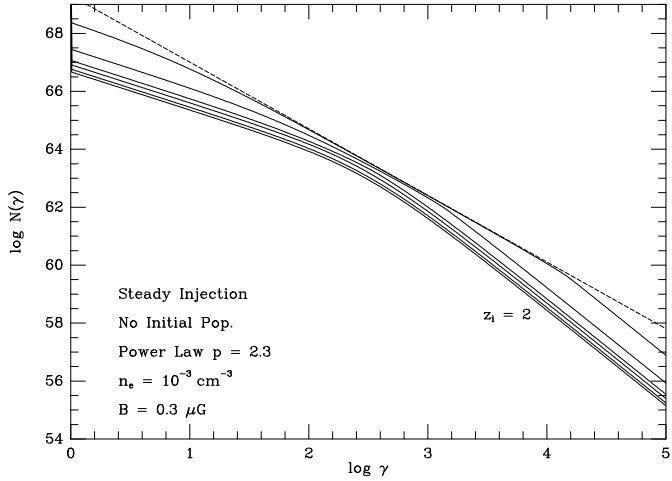}}}
\caption[]{The present day relativistic electron populations in a series of
models with ongoing particle acceleration, perhaps due to a cluster merger
shock.
he solid curves show models for clusters which started at redshifts of
$z_i = 2$, 1, 0.5, 0.3, 0.1, and 0.01 (bottom to top).
The short-dashed curve gives the total power-law spectrum of all of the
injected particle over the cluster lifetime.
\label{fig:model_steady}}
\end{figure}

Fig.\,\ref{fig:model_steady} shows the resulting present-day electron spectra
in models in which there is a current source of accelerated electrons,
such as a cluster merger shock.
Again, Coulomb losses reduce and flatten the electron population at low
energies, and IC and synchrotron losses reduce and steepen the electron
population at high energies.
However, these losses don't remove the high energy particles completely,
as new particles are always being accelerated.
Instead, the population quickly relaxes into steady-state, where the
low (high) energy population is one power flatter (steeper) than the
injected electrons
(Ginzburg \& Syrovatskii 1964).

\begin{figure}
\resizebox{\hsize}{!}{\rotatebox{-90}{\includegraphics{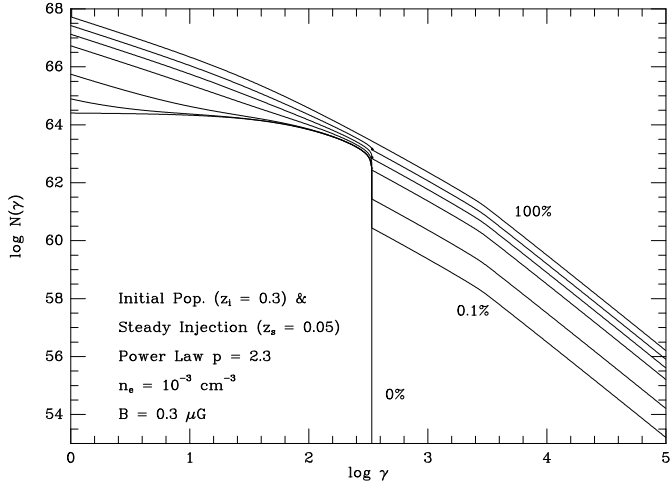}}}
\caption[]{The electron energy spectra in models with both an initial
electron population and current injection of new particles, perhaps
due to an ongoing cluster merger.
The values of the fraction of particle energy due to current injection
are (top to bottom)
$F_{inj} = 100$\%, 50\%, 25\%, 10\%, 1\%, 0.1\%, and 0\%.
\label{fig:model_both}}
\end{figure}

Finally, we consider models in which there is an initial population in
the cluster, and in which there also is current injection of
new particles.
These might apply to the more realistic cases where the cluster had
particle acceleration associated with the formation of the cluster and
mergers in the past, but also has a subcluster merger occurring at present.
An illustrative set of such models are shown in
Fig.\,\ref{fig:model_both}.
The different models are characterized by differing values of $F_{inj}$,
which is the fraction of the particle energy which has been injected by
the current steady injection process, as opposed to the earlier
population.
If the electrons were injected in cluster shocks, then $F_{inj}$ might
be roughly proportional to the fraction of the thermal energy content
of the ICM which is being generated in the current shock.
The general features of these models are that the electron spectrum is
moderately flat below $\gamma \approx 300$.
Unless there is a very strong present merger or other current particle source,
the electron population drops off rapidly above $\gamma \approx 300$.
In most models for clusters, most of the relativistic electron energy should
be in particles with $\gamma \approx 300$;
these are the particles with the longest lifetimes
(Fig.\,\ref{fig:lifetime}).
If a cluster has a ongoing merger or other particle source, then there
will also be a high energy tail to the particle distribution.
Because this tail is likely to be in steady-state, the electron
distribution is likely to be one power steeper than the source,
perhaps $N( \gamma ) \propto \gamma^{-3.3}$.

\section{Inverse Compton Emission} \label{sec:ic}

The emission produced by the relativistic electron populations through the
inverse Compton scattering of CMB photons was also calculated.
Let $L_\nu d \nu$ be the luminosity of IC emission at frequencies from
$\nu$ to $\nu + d \nu$.
Then, the spectrum of IC emission is related to the electron spectrum
$ N ( \gamma )$ by
\begin{equation} \label{eq:ic}
L_\nu = 12 \pi \sigma_T \,
\int_1^\infty N ( \gamma ) \, d \gamma \,
\int_0^1 J \left( \frac{\nu}{4 \gamma^2 x } \right) \, {\cal F} ( x ) \, dx \, ,
\end{equation}
where
${\cal F} ( x ) \equiv 1 + x + 2 x \ln x - 2 x^2$.
Here, $J ( \nu )$ is the mean intensity at a frequency $\nu$ of the
radiation field being scattered.
For the CMB, this is just the black body function
$J ( \nu ) = B_\nu ( T_{CMB} )$.
For a power-law electron spectrum $N( \gamma ) \propto \gamma^{-p}$,
the IC spectrum is a power-law
$L_\nu \propto \nu^\alpha$, where $\alpha = - ( p - 1 ) / 2$.
This gives $\alpha = -0.65$ for the injection spectrum of $p = 2.3$.

Fig.\,\ref{fig:ic} shows the resulting IC spectra for a variety of
different models; for details of the specific models, see Sarazin (1999).
Model 1 has a current source of electrons; for example, this
might be due to a subcluster merger.
At frequencies above $10^{17}$ Hz, the spectrum quickly steepens with time
due to IC losses, and becomes a half power steeper than that given by
the injected electron spectrum,
$\alpha \approx - ( p / 2 ) \approx -1.15$.
At low frequencies, the spectrum flattens more gradually due to Coulomb
losses, approaching a spectrum which is a half power flatter than that
due to the injected spectrum,
$\alpha \approx - ( p - 2 ) / 2 \approx -0.15$.
If the current acceleration process has occurred since a redshift of
$ z \ga 0.2$,
the spectrum is near that of a steady-state electron population, with
these two power-laws meeting at a knee at $\nu \sim 3 \times 10^{16}$ Hz.
If the current acceleration process is more recent, there is a central
region of the IC spectrum where the electrons are not yet in steady-state.
Here, the spectral index is $\alpha \approx -0.65$.

Models 11 and 22 have no current source of electrons (e.g., no ongoing
cluster merger).
The electron energy distributions of these models have a cut off at high
energies, $\gamma_{max}$, which results from
rapid IC and synchrotron losses by high energy electrons
(see Fig.\,\ref{fig:model_init}).
As a result, the IC spectra have a very rapid fall off at high
energies (exponentially or slightly faster; see eq.~[54] in Sarazin
[1999]).

At low frequencies, the IC spectra of models without current acceleration
flatten and become slowly increasing with frequency as the models
age.
We expect that the low energy electron spectrum in an older model without
current acceleration will become nearly independent of
$\gamma$ except for a slowly varying logarithmic factor
(see eq.~36 in Sarazin [1999]).
If $N( \gamma ) \approx N_{low} = $ constant at low energies,
then the low frequency IC spectrum varies as $L_\nu \propto \nu^{1/2}$.

Model 27 in Fig.\,\ref{fig:ic} shows the spectrum of a model with
both an initial population of particles
and electron acceleration at the present time.
In this model, the fraction of electron being contributed by the
current acceleration (e.g., current merger shock) is $F_{inj} = 1$\%.
Models in which the current rate of particle injection provides a
small but significant fraction of the total electron energy have
hybrid spectra.
At low frequencies ($\nu \la 10^{17}$ Hz), they have an extended hump of
emission, with a rapid fall off above $\nu \sim 10^{16}$ Hz.
However, they also have an extended hard tail of emission at high
frequencies, which has a power-law spectrum with a spectral index of
$\alpha \approx - ( p / 2 ) \approx -1.15$.
At the bottom of Fig.\,\ref{fig:ic}, there is a scale which shows
the portion of the electromagnetic spectrum involved.

\subsection{EUV and Soft X-ray Emission} \label{sec:ic_euv}

\begin{figure}
\resizebox{\hsize}{!}{\rotatebox{-90}{\includegraphics{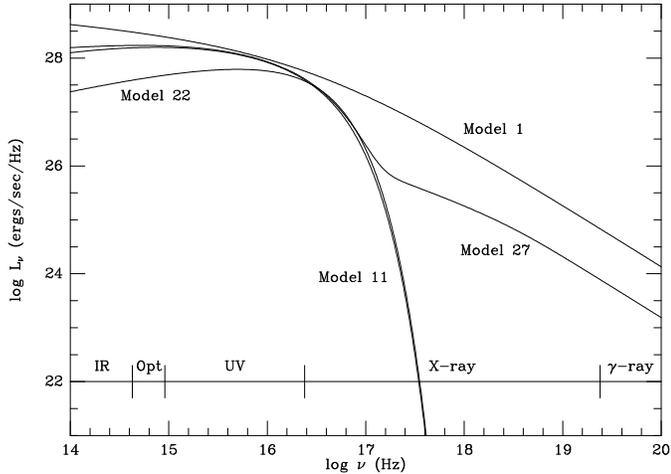}}}
\caption[]{The IC emission spectra for different types of cluster models
(for details on the models, see Sarazin [1999]).
Models 1 has current electron acceleration (perhaps due to an ongoing
merger shock), but no initial electron population
(see Fig.\,\protect\ref{fig:model_steady}).
Model 11 has an initial electron population (perhaps due to the formation
of the cluster or earlier mergers), but no current electron acceleration
(see Fig.\,\protect\ref{fig:model_init}).
Model 22 is identical to Model 11, but the initial electron energy
spectrum has a break at an energy of 1 GeV.
Model 27 has both an initial electron population, and a source of
electrons at present
(see Fig.\,\protect\ref{fig:model_both}).
The scale near the bottom of the Figure shows the portion of the
electromagnetic spectrum.
\label{fig:ic}}
\end{figure}

Fig.\,\ref{fig:ic} shows that the fluxes from different models all tend
to agree at frequencies $\sim$$2 \times 10^{16}$ Hz, which corresponds
to a photon energy of $\sim$80 eV.
Thus, EUV and very soft X-ray emission might be expected to be a universal
property of clusters of galaxies, if they have all produced populations
of relativistic electrons with total energies which are at least
$\sim$$10^{-2}$ of their thermal energies.
The one constraint is that at least a portion of these particles
must have been injected at moderate to low redshifts, $z \la 1$.
The fact that essentially all of the models produce a strong and
similar flux at EUV energies may help to explain the fact the excess
EUV emission has been detected in all of the cluster observed with the
{\it Extreme Ultraviolet Explorer} ({\it EUVE})
satellite and which lie in directions of sufficiently low Galactic
columns that this radiation is observable
(Lieu et al.\ 1996a,b;
Mittaz, Lieu, \& Lockman 1998;
Sarazin \& Lieu 1998).
Values of $L_\nu (EUV)$ at a frequency of $v = 2 \times 10^{16}$ Hz
are listed in Table~3 of Sarazin (1999).
From these values, one finds that the emission at EUV energies is
fairly directly related to 
$E^{tot}_{CR,e}$, the total amount of energy injected in electrons
with $\gamma \ge 300$.
The average relationship is about
\begin{equation} \label{eq:euv_lum}
L_\nu (EUV) \sim 6 \times 10^{27} \,
\left( \frac{E^{tot}_{CR,e}}{10^{63} \, \rm{ergs}} \right) \, {\rm ergs}
\, .
\end{equation}
Models which are likely to apply to real clusters
agree with equation~(\ref{eq:euv_lum}) to within a factor of 4.

While the amount of EUV emission is fairly constant from model to model,
the spectrum depends strongly on the amount of recent particle injection.
Very steeply declining spectra occur in models without current particle
acceleration and with small values of $\gamma_{max}$.
For many of the models, the spectral can be fit by a power-law with
$-1.5 \ga \alpha_{EUV} \ga -0.6$,
but very large negative values also occur.
The spectrum drops more rapidly with frequency as the proportion of
electrons which are currently being accelerated decreases.

Because the {\it EUVE} observations of clusters have no spectral resolution,
there is no detailed information on the observed spectrum of EUV 
emission
(Lieu et al.\ 1996a,b;
Mittaz, Lieu, \& Lockman 1998).
However, the ratio of {\it EUVE} fluxes to those in the softest bands of the
{\it ROSAT} PSPC suggest that the EUV spectra are generally steeply
declining.
The spectra of the Coma cluster is less steeply declining than that of
Abell 1795;
this might be understood as the result of particle acceleration by merger
shocks, since Coma appears to be undergoing a merger or mergers
(e.g., Burns et al.\ 1994),
while Abell 1795 appears regular and relaxed
(e.g., Briel \& Henry 1996).
Also, the EUV spectra appear to get steeper with increasing radius in
the Abell 1795 cluster
(Mittaz, Lieu, \& Lockman 1998).
This might be the result of decreasing gas density with radius, which
will decrease the effect of Coulomb loses at low energies
(Fig.\,\ref{fig:lifetime}).

\subsection{Hard X-ray IC Emission and Radio Halos} \label{sec:ic_hxr}

At photon energies of 0.5 to 10 keV,
the dominant emission in most clusters is the thermal emission from the
hot ICM, and IC emission will be difficult to detect.
However, at energies $\ga$20 keV, IC emission should again become
observable, assuming that it is dropping as power-law function of
frequency, while the thermal emission drops as an exponential.
IC emission at photon energies of $\sim$50 keV will be produced by
electrons with $\gamma \sim 10^4$.
These particles have rather short lifetimes ($t_{loss} \ll 10^9$ yr),
and are only present in clusters in which there has been substantial
electron acceleration since $z \la 0.03$.
As Fig.\,\ref{fig:ic} shows, only the models with current or very recent
particle acceleration have any significant HXR emission.

Because of the short lifetimes of the particles producing HXR emission,
these electrons are likely to be close to steady-state if present in
significant quantities.
The expected steady-state spectral index if IC losses dominated would be
$\alpha_{HXR} = - ( p + 1 ) / 2 \approx -1.65$.
The best-fit spectral indices are flatter than this,
$\alpha_{HXR} \approx -1.1$, mainly because other loss processes are
important at the lower energy end of the HXR band
(Fig.\,\ref{fig:losses}).
All of the HXR spectral shapes in the models are very similar, as expected
for steady-state populations.
The differences in the HXR luminosities just reflect differences in
the present rate of electron acceleration.
To a good approximation, the present day value of the hard X-ray IC
luminosity $L_{HXR}$
(20--100 keV) is simply given by
\begin{equation} \label{eq:hxr_lum}
L_{HXR} \approx 0.17 {\dot{E}}_{CR,e} ( \gamma > 5000 ) \, .
\end{equation}
where ${\dot{E}}_{CR,e} ( \gamma > 5000 )$ is the total present rate of
injection of energy in cosmic ray electrons with $\gamma > 5000$.
The best-fit coefficient (0.17 in eqn.~\ref{eq:hxr_lum}) depends somewhat
on the power-law index of the injected electrons; the value of 0.17 applies
for $p = 2.3$.

The same relativistic electrons which produce HXR emission by IC scattering
will also produce radio emission by synchrotron emission.
The synchrotron radio emission from these models is given in
Sarazin (1999).
Electrons with a Lorentz factor of $\gamma$ produce radio emission
with $\nu \sim 100 ( B / \mu G) ( \gamma / 10^4 )^2 $ MHz, and rather
high electron energies ($\gamma \ga 10^4$) are needed to produce
observable radio emission.
Thus, these are likely to be the same electrons which produce the hardest
HXR emission.
In general, HXR emission and radio synchrotron emission are expected only
in clusters with very recent or current acceleration of relativistic
electrons.
Thus, both measure the current rate of particle injection.
For example, if the particles are accelerated in ICM shocks, HXR and
radio emission would be expected only in clusters which are currently
undergoing (or which very recently underwent) a merger.

Recently, excess hard X-ray emission has been detected in the Coma
(Fusco-Femiano et al.\ 1999)
and Abell~2199
(Kaastra, Bleeker, \& Mewe 1998)
clusters with $BeppoSAX$.
It is tempting to attribute this HXR radiation
to IC emission from higher energy electrons.
Of course, Coma has a radio halo and abundant evidence for merger
activity.
Thus, it is consistent with this hypothesis, although the required
magnetic field ($B = 0.15$ $\mu$G) is smaller than might have been
expected.
On the other hand, Abell~2199 is a relaxed cluster with a strong cooling
flow.
It has no radio halo emission, with an upper limit on the radio flux which
is inconsistent with the IC interpretation of the HXR emission unless
$B \la 0.01$ $\mu$G
(Kempner \& Sarazin 1999).
Thus, it seems unlikely that the HXR in Abell~2199 is due to IC scattering
of CMB photons.
It is possible that part of the HXR emission may be due to nonthermal
bremsstrahlung (bremsstrahlung from superthermal but not strongly
relativistic electrons).

\subsection{Optical and UV Emission} \label{sec:ic_optuv}

One also expects that the lower energy portion of the cosmic ray population
in clusters will produce diffuse optical and UV emission.
Diffuse optical emission is known to exist in many clusters, particularly
in those with central cD galaxies
(e.g., Boughn \& Uson 1997).
Although the origin is not completely understood, the optical colors
of the diffuse light suggest that it is due to old stars, which may have
been stripped from cluster galaxies.
It is likely that the near or vacuum UV are better regions to detect
low surface brightness diffuse emission due to IC emission,
since the older stellar population in E and S0 galaxies (and, presumably,
the intracluster stellar population) are fainter there.

The IC optical-UV emission from the models is given in Sarazin (1999).
The predicted fluxes are rather low when compared to the diffuse optical
emission seen in clusters of galaxies or to the sensitivity of current
and UV instruments for detecting diffuse emission.
The UV powers of most of the models lie within a relatively narrow range
of about an order of magnitude.
Unless the electron density in the ICM is much higher than
0.001 cm$^{-3}$ and Coulomb losses are catastrophic or the electron
population in clusters is very old, one would always expect a significant
population of lower energy electrons.
The IC spectra in the optical and UV are fairly flat, with spectral indices
of $-0.5 \ga \alpha_{Opt,UV} \ga 0.3$.
In models with current electron acceleration, the lower energy electron
population should be approaching steady-state with the Coulomb
losses, and the expected power-law for the IC emission is
$\alpha \approx - ( p - 2 ) / 2 \approx -0.15$.
For models without current acceleration, the electron
spectrum is expected to be nearly flat at low energies, and the
$L_\nu \sim \nu^{1/2}$.
Thus, the spectral index should be slightly positive in these models.

\section{Gamma-Ray Emission} \label{sec:gamma}

\begin{figure}
\resizebox{\hsize}{!}{\rotatebox{-90}{\includegraphics{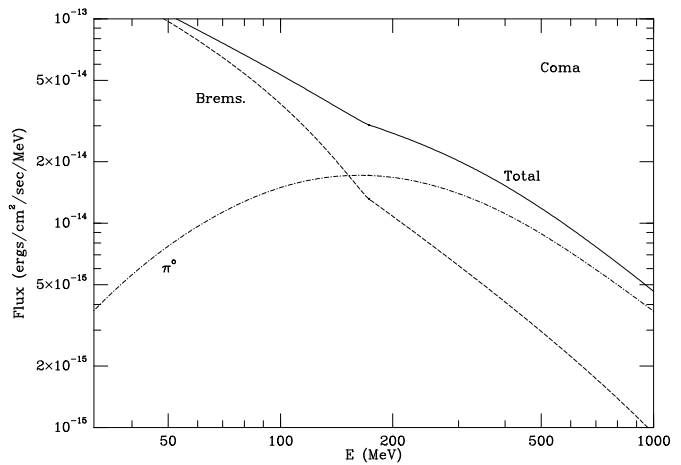}}}
\caption[]{The predicted gamma-ray spectrum of the Coma cluster in the
region around 100 MeV.
The emission is mainly the result of bremsstrahlung by relativistic
electrons and $\pi^o$ decay due to relativistic ions.
The electron population was determined by the best-fit {\it EUVE} fluxes and
spectra.
The ions have 10 time the electron energy.
\label{fig:gamma}}
\end{figure}

The same relativistic elections and ions will produce gamma-ray
emission.
This involves a wide variety of physical processes;
in the interest of brevity, only the region around 100 MeV will
be discussed here.
The two major emission processes in this region of the gamma-ray
spectrum are electron bremsstrahlung (due to collisions between
relativistic electrons and thermal ions and electrons), and $\pi^o$ decay,
due to collisions between relativistic ions and thermal ions).
Fig.\,\ref{fig:gamma} shows the predicted spectrum for the Coma cluster.
The electron population was determined by the IC emission required to
fit the observed {\it EUVE} flux and {\it EUVE/ROSAT} flux ratio
(Lieu et al.\ 1996a).
The ratio of the $\pi^o$ decay component to the bremsstrahlung component
is determined by the ratio of relativistic ions to electrons.
In Fig.\,\ref{fig:gamma}, this ratio was taken to be 10.

The predicted flux of gamma-rays with $h \nu \ge 100$ MeV is about
$2 \times 10^{-8}$ cts/cm$^2$/s.
For comparison, the {\it EGRET} upper limit on the flux from Coma in
this range is
$\le 4 \times 10^{-8}$ cts/cm$^2$/s
(Sreekumar et al.\ 1996).
Thus, it seems certain that {\it GLAST} will detect Coma and many other
clusters, assuming that the detected EUV emission from clusters is
due to IC from relativistic electrons.
The measurement of the gamma-ray spectrum in the region around 100 MeV
will allow the ratio of relativistic ions to electrons to be determined.
In many astrophysical environments, this is an unknown quantity.
The {\it EGRET} limit already implies that this ratio is $\la 30$ in
the Coma cluster.

\begin{acknowledgements}
This work was supported in part by NASA Astrophysical Theory Program grant
NAG 5-3057.
I would like to thank Hans B\"ohringer and Luigina Feretti for organizing
such an enjoyable and useful workshop.
\end{acknowledgements}

\end{document}